\documentstyle[aps]{revtex}
\title{Curvature singularity of the distributional BTZ black hole geometry}
\author{ N.R. Pantoja\footnotemark[1],
    H. Rago\footnotemark[2]
    and R.O. Rodr\'{\i}guez\footnotemark[3]}
\address{    \footnotemark[1]{Centro de Astrof\'{\i}sica Te\'{o}rica, Universidad de Los Andes,
    M\'{e}rida, 5101, Venezuela.}\\
    \footnotemark[2]{Laboratorio de F\'{\i}sica Te\'{o}rica and Centro de Astrof\'{\i}sica Te\'{o}rica, Universidad de Los Andes,
    M\'{e}rida, 5101, Venezuela.}\\
    \footnotemark[3]{Departamento de F\'{\i}sica, Universidad de Los Andes,
    M\'{e}rida, 5101, Venezuela.}}
\begin{document}
\maketitle
    \newcommand{\ttensor}[2]{\mbox{$\left(\parbox{3ex}{\centerline{#1}\par \centerline{#2}}\right)$}}

\begin{abstract}
For the non-rotating BTZ black hole, the distributional curvature
tensor field is found. It is shown to have singular parts
proportional to a $\delta$-distribution with support at the
origin. This singularity is related, through Einstein field
equations, to a point source. Coordinate invariance and
independence on the choice of differentiable structure of the
results are addressed.
\end{abstract}

\pacs{PACS numbers: 0250N, 9760L}

\section{Introduction}

The $(2+1)$-dimensional black hole of Ba\~{n}ados, Teitelboim and
Zanelli (BTZ)\cite{banadosPRL,banadosPRD} provides us with a
useful model to study various classical and quantum aspects of
black hole physics\cite{carlip}. The BTZ black hole shares many of
the properties of the more complicated $3+1$-dimensional Kerr
black hole. However, it differs from the Kerr solution in which it
is asymptotically anti-de Sitter rather than asymptotically flat.
Furthermore, by admitting closed time-like curves, the rotating
BTZ black hole has no curvature singularities. Nevertheless, when
there is no angular momentum the spacetime fails to be Hausdorff
at the origin and turns out to be singular\cite{banadosPRD}.

The purpose of this work is to analyse the distributional BTZ
black hole geometry for the non-rotating case. There are several
reasons, both mathematical and physical, to carry out such
analysis. The non-rotating BTZ black hole provides an example of a
singular spacetime whose singularities can not be identified with
the unboundedness of some scalar constructed from the curvature
tensor. In this sense, it resembles the well known conical
singularities\cite{deser84-1} whose meaning has attracted wide
interest for many years\cite{'t hooft}. On the other hand,
regularization procedures required to multiply distributions need
not be invoked in the calculation of the distributional
non-rotating BTZ black hole curvature tensor. We show that the
non-rotating BTZ black hole metric belongs to the class of
semi-regular metrics, as defined in Ref. \cite{garfinkle}, for
which the curvature tensor field has direct distributional
meaning. Note that not many semi-regular metrics are known. It was
shown that the $(3+1)$-dimensional Minkowski metric with an
angular deficit, and a certain kind of traveling wave metric are
semi-regular \cite{garfinkle}. Recently, the metric associated to
the $(2+1)$-dimensional spacetime around a point source
\cite{deser84-1} was demonstrated to be a third example
\cite{pantoja}.

The distributional description of the BTZ black hole spacetime is
carried out in Schwarzschild coordinates and in Kerr-Schild
coordinates. We find the distributional Ricci and Einstein tensor
fields for the non-rotating BTZ black hole, which turn out to be
equivalent in both coordinate systems. This indicates that,
although the intermediate calculations depend upon the choice of
coordinates, the final results does not. Remarkably, a complete
agreement with what is physically expected is found. The
distributional curvature tensor becomes, besides the constant
curvature part, a $\delta$-distribution supported at the origin.
Furthermore, this singularity is related through Einstein field
equations to a point source.

The paper is organized as follows. In the next section, following
the procedure of \cite{garfinkle}, the distributional curvature
and Einstein tensor fields are found for the non-rotating BTZ
solution in Schwarzschild coordinates. In section \ref{sec3}, we
carry out the distributional analysis using the Kerr-Schild form
of the BTZ solution. The last section is devoted to summarize and
discuss the coordinate invariance (and differentiable structure
dependence) of our results.

\section{BTZ black hole in Schwarzschild
coordinates}\label{sec2}

The non-rotating BTZ black hole solution
\cite{banadosPRL,banadosPRD} written in Schwarzschild coordinates
is given by
\begin{equation}
g_{ab}= -(-m+\frac{r^2}{l^2})dt_adt_b +
(-m+\frac{r^2}{l^2})^{-1}dr_adr_b + r^2d\varphi_a
d\varphi_b,\label{btzsc}
\end{equation}
where $-\infty<t<\infty$, $0<r<\infty$ and $0\leq \varphi <2\pi$,
with the surfaces $\varphi=0,2\pi$ identified. The dimensionless
quantity $m$ is the mass parameter. For this metric we have
\begin{equation}
R_{abc}^{\quad d}= g_{ac}R_b^{\ d}-g_{bc}R_a^{\ d}+ \delta_b^{\
d}R_{ac}-\delta_a^{\ d}R_{bc}-\frac{1}{2}(g_{ac}\delta_b^{\
d}-g_{bc}\delta_a^{\ d})R,
\end{equation}
where
\begin{equation}
R_{ab}= -\frac{2}{l^2}g_{ab}.
\end{equation}
Hence, (\ref{btzsc}) has constant negative curvature.

The BTZ metric (\ref{btzsc}) is a solution of the vacuum Einstein
field equations with cosmological constant $\Lambda=-{1}/{l^2}$,
\begin{equation}
R_{ab} - \frac{1}{2}g_{ab}R + \Lambda g_{ab} =0
\end{equation}
and may be obtained by identifying certain points of (the covering
manifold of) the anti-de Sitter space
\cite{banadosPRL,banadosPRD}. For $m>0$, (\ref{btzsc}) describes a
black hole of mass $m$ with horizon at $r_+ = \sqrt{m} l$. Note
that the black hole is characterized, in the present context, by
the existence of an event horizon and not by the existence of a
region of large curvature. The solution with $-1< m <0$ may be
associated to the metric generated by a point source at the origin
\cite{deser84-2}. The solution with $m=-1$ is anti-de Sitter
space. The massless black hole, $m=0$, is commonly considered as
the vacuum state. For a review of the properties of BTZ black
holes see \cite{carlip}.

In (\ref{btzsc}), $\sqrt{g_{\varphi \varphi}}= r$ represents the
radius associated with the proper circumference. Therefore with
$x=r \cos\varphi$ and $y=r\sin \varphi$ we have
\begin{equation}
g_{ab}= \eta_{ab} + (1+m-\frac{r^2}{l^2})dt_a dt_b -
\left(\frac{l^2+ml^2-r^2}{ml^2-r^2}\right)\frac{1}{r^2}(xdx_a +
ydy_a)(xdx_b + ydy_b),\label{sc}
\end{equation}
where $r=\sqrt{x^2 + y^2}$ and $\eta_{ab}$ is the ordinary
Minkowski metric on ${\cal R}^3$. The metric (\ref{sc}) is
singular when $r=0$, except for $m=-1$. This singularity is not a
coordinate singularity. The spinless BTZ black hole is
geodesically incomplete \cite{cruz} and the singularity at $r=0$
corresponds to fixed points of the identifications from which the
BTZ solution is obtained \cite{steif}. Assuming that (\ref{sc})
can be extended to $r=0$, we look for the distributional curvature
tensor and its relation with a possible distributional source.

Suppose $({\cal M},g_{ab})$ are given such that
\begin{enumerate}
\item $g_{ab}$ and $(g^{-1})^{ab}$ exist almost everywhere and are
locally integrable,
\item the weak first derivative $\nabla_c g_{ab}$ of $g_{ab}$ in a
smooth derivative operator $\nabla_c$ exists and the tensors
\begin{equation}
C^c_{ab} \equiv \frac{1}{2}(g^{-1})^{cd}(\nabla_a g_{bd} +
\nabla_b g_{ad} - \nabla_d g_{ab}) \label{chris},
\end{equation}
and $C^d_{m[b}C^m_{a]c}$ are locally integrable.
\end{enumerate}
These are the minimal conditions for $R_{abc}^{\quad d}$ to be
definable as a distribution by the usual coordinate formula,
\begin{equation}
R_{abc}^{\quad d}= {\tilde{R}}_{abc}^{\quad d} +
2\nabla_{[b}C^d_{a]c} + 2 C^d_{m[b}C^m_{a]c},\label{RR}
\end{equation}
where ${\tilde{R}}_{abc}^{\quad d}$ is the curvature tensor
associated to the smooth derivative operator
$\nabla_c$\cite{garfinkle}. We shall say that $g_{ab}$ is a
semi-regular metric.

A semi-regular metric may have no distributional Einstein tensor
due to the fact that contractions of the metric with the curvature
tensor may have no sense as distributions. Stronger conditions can
be imposed to isolate the class of metrics for which the
distributional meaning of the Einstein tensor is ensured, but then
the distributional curvature tensor must have its support on a
submanifold of codimension of at most one \cite{geroch}. Metrics
for surface layers \cite{israel} lie in this class, but neither
strings nor point particles can be described by metrics in this
class. Alternatively, by considering Colombeau's generalized
functions\cite{colombeau}, distributional curvatures can be
defined for those cases where a direct calculation would not work.
This approach has been used to obtain the distributional curvature
associated with a conical singularity \cite{clarke,wilson,vickers}
and to the analysis of the distributional Schwarzschild geometry
\cite{heinzle}, but we will not consider it here.

We shall prove that (\ref{sc}) is a semi-regular metric. We take
for the differentiable structure that in which $t,x$ and $y$ form
a smooth chart. For a test tensor field $U^{ab}$ on ${\cal R}^3$
we have
\begin{equation}
g_{ab}[U^{ab}]\equiv \int_{{\cal R}^3}g_{ab}U^{ab}\omega_{\eta} =
\int_{{\cal R}^3}\eta_{ab}U^{ab}\omega_{\eta} + \int_{{\cal
R}^3}(1+m-\frac{r^2}{l^2})U^{tt}\omega_{\eta} - \int_{{\cal
R}^3}\left(\frac{l^2+ml^2-r^2}{ml^2-r^2}\right)dr_adr_bU^{ab}\omega_{\eta},
\label{gab1}
\end{equation}
where $\omega_{\eta}$ is the volume element associated to
$\eta_{ab}$ and it is understood that all tensor components are
Cartesian components as functions of Cartesian coordinates. For
$m>0$ we require that $U^{ab}$ be a test tensor with support on
$r<\sqrt{m}l$, while for $m<0$ we simply require that $U^{ab}$ be
a test tensor of compact support. The black vacuum can not be
handled in the Schwarzschild-type coordinates because the last
term in the right-hand side of (\ref{gab1}) is not locally
integrable for $m=0$. Note that, with this choice, the natural
volume element $\omega_{g}$ associated to $g_{ab}$ agrees with the
volume element $\omega_{\eta}$ of $\eta_{ab}$.

It follows that
\begin{eqnarray}
    g_{ab}[U^{ab}]& = &
    \int_{{\cal R}^3}\eta_{ab}U^{ab}\omega_{\eta} +
\int_{{\cal R}^3}(1+m-\frac{r^2}{l^2})U^{tt}\omega_{\eta}
\nonumber\\
    &&- \int_{{\cal R}^3}\left(\frac{l^2+ml^2-r^2}{ml^2-r^2}\right)
    (\cos^2\varphi U^{xx} +
    \cos \varphi \sin \varphi (U^{xy} + U^{yx}) + \sin^2
    \varphi U^{yy}) \omega_{\eta}. \label{gab2}
\end{eqnarray}
Therefore, $g_{ab}$ is locally integrable for $m\neq 0$.

Next, let $U_{ab}$ be a test tensor field on ${\cal R}^3$. For
\begin{equation}\label{gab-1}
(g^{-1})^{ab} \equiv \eta^{ab} +
\left(\frac{l^2+ml^2-r^2}{ml^2-r^2}\right)\partial_t^a\partial_t^b
-(1+m-\frac{r^2}{l^2}) \partial_r^a\partial_r^b,
\end{equation}
we have
\begin{eqnarray}
(g^{-1})^{ab}[U_{ab}] &=& \int_{{\cal R}^3} \eta^{ab}U_{ab}
\omega_\eta + \int_{{\cal R}^3}
\left(\frac{l^2+ml^2-r^2}{ml^2-r^2}\right) \ U_{tt} \
\omega_\eta\nonumber\\ &&  -\int_{{\cal R}^3}\left(
1+m-\frac{r^2}{l^2}\right) (\cos^2\varphi U_{xx} + \cos \varphi
\sin \varphi (U_{xy} + U_{yx}) + \sin^2 \varphi U_{yy}) \
\omega_{\eta}.
\end{eqnarray}
Therefore, $(g^{-1})^{ab}$ is locally integrable for $m\neq 0$.

We now calculate the weak derivative in $\eta_{ab}$ of $g_{ab}$.
Let $U^{abc}$ be a test tensor field. We find
\begin{equation}
 \nabla_c g_{ab} [U^{cab}] \equiv - \int_{{\cal R}^3}
 g_{ab}\nabla_c U^{cab} \omega_\eta = \int_{{\cal R}^3} W_{cab}
 U^{cab} \omega_\eta,
\end{equation}
where $W_{cab}$ is the locally integrable but not locally square
integrable tensor given by
\begin{equation}
  W_{cab} = -\frac{2r}{l^2} dr_c (dt_adt_b +
  (m-\frac{r^2}{l^2})^{-2} dr_a dr_b) -
  \frac{1}{r}\left(\frac{l^2+ml^2-r^2}{ml^2-r^2}\right)
  rd\varphi_c (rd\varphi_a dr_b + rdr_a d\varphi_b).
\end{equation}

From (\ref{chris}), we find
\begin{eqnarray}\label{chrissc}
  C^{c}_{ab} &=& \frac{-r (ml^2-r^2)}{l^4} dt_a dt_b \partial_r^c -
  \frac{r}{ml^2-r^2} (dr_a dt_b +dt_a dr_b) \partial_t^c
  \nonumber\\
   && + \frac{r}{ml^2-r^2} dr_a dr_b \partial_r^c + \frac{1}{r}
  \frac{ml^2 +l^2 -r^2}{l^2} r^2 d\varphi_a d\varphi_b
  \partial_r^c,
\end{eqnarray}
which is locally integrable. On the other hand,
\begin{equation}\label{doschrissc}
  2C^d_{m[b}C^m_{a]c} = \frac{2r}{ml^2-r^2} (dt_a dr_b -dr_a dt_b)
  \left( \frac{r}{ml^2-r^2} dr_c \partial_t^d - \frac{r(ml^2-r^2)}{l^4}
  dt_c \partial_r^d \right),
\end{equation}
which is locally integrable. Hence, $g_{ab}$ is a semi-regular
metric in the differentiable structure chosen.

Now, contracting (\ref{RR}) and using
(\ref{chrissc},\ref{doschrissc}) we find for the Ricci tensor of
$g_{ab}$
\begin{equation}\label{riccisc}
  R_{ac}[U^{ac}] = - \int_{{\cal R}^3} C^b_{ac}\nabla_b U^{ac}
  \omega_\eta - \int_{{\cal R}^3} C^b_{ma} C^m_{bc} U^{ac}
  \omega_\eta,
\end{equation}
where
\begin{equation}\label{derivative}
 - \int_{{\cal R}^3} C^b_{ac}\nabla_b U^{ac} \omega_\eta = -
 \int_{r=\varepsilon} dr_b C^b_{ac} U^{ac} \sigma +
 \int_{r>\varepsilon} \nabla_b C^b_{ac} U^{ac} \omega_\eta,
\end{equation}
with
\begin{equation}\label{deriv}
  \nabla_b C^b_{ac} =  - \frac{2(ml^2-2r^2)}{l^4} dt_a dt_c  +
  2\frac{ml^2}{(ml^2-r^2)^2} dr_a dr_c - 2\frac{r^2}{l^2}
  d\varphi_a d\varphi_c,
\end{equation}
which is a locally integrable tensor and where it is understood
that $\varepsilon\rightarrow 0$.

From (\ref{riccisc}) and
(\ref{doschrissc},\ref{derivative},\ref{deriv}) we find
\begin{equation}
  R_{ac}[U^{ac}] = \pi (1+m) \int dt \left( U^{xx}(t,\vec{0}) +
  U^{yy}(t,\vec{0}) \right) - \frac{2}{l^2} \int_{{\cal R}^3}
  g_{ac} U^{ac} \omega_\eta,
\end{equation}
where $g_{ac}$ is the locally integrable tensor defined by
(\ref{gab2}). Thus we obtain
\begin{equation}\label{ricdis}
  R_{ac} = \pi(1+m) \delta^{(2)}_{(0)} (dx_a dx_c + dy_a dy_c) -
  \frac{2}{l^2} g_{ac}.
\end{equation}
Note that $\forall m>-1$, the Ricci tensor has a singular part
proportional to a $\delta$ distribution. As expected, for $m=-1$
the singular part of the curvature is absent, as follows from the
fact that in this case we have $AdS_3$ spacetime. Since for
$-1<m<0$, there is no horizon, we have a naked singularity at
$r=0$. For $m>0$ we have a singularity at $r=0$ hidden by a
horizon at $r_+ = \sqrt{m}l$. As stated before, for $m=0$ the
metric (\ref{gab1}) is not a semi-regular metric and the massless
BTZ black hole can not be properly discussed from the above
derivation.

We now calculate the Einstein tensor of $g_{ab}$. Define
\begin{equation}
G^{a}_{\, b}= R^{a}_{\, b} -
\frac{1}{2}(g^{-1})^{cd}{\tilde{R}}_{cd}\delta^{a}_{\, b} +
(g^{-1})^{cd}C^e_{m[c}C^m_{e]d}\delta^{a}_{\, b} +
\nabla_{[c}\left(C^e_{e]d}(g^{-1})^{cd}\right)\delta^{a}_{\, b} +
C^e_{d[c}\nabla_{e]}(g^{-1})^{cd}\delta^{a}_{\, b},\label{ERM}
\end{equation}
where
\begin{equation}
R^{a}_{\, b}= (g^{-1})^{ac}{\tilde{R}}_{cb} + 2\nabla_{[c}\left(
C^{c}_{d]b}(g^{-1})^{ad}\right) +
2C^{c}_{b[c}\nabla_{d]}(g^{-1})^{ad} +
2(g^{-1})^{ad}C^{c}_{m[c}C^{m}_{d]b}.\label{RicciM}
\end{equation}
We shall say that (\ref{ERM}) is the Einstein tensor distribution
of (\ref{sc}), whenever each term in the right-hand sides of
(\ref{ERM},\ref{RicciM}) defines a distribution.

From (\ref{chrissc},\ref{doschrissc}) and (\ref{RicciM}) we find
\begin{eqnarray}\label{Rdc}
  R^d_{\,c} &=& \nabla_b \left((g^{-1})^{ad} C^b_{ac}\right)\nonumber\\
  &=& \nabla_b
  \left( -\frac{r}{l^2}\partial_t^ddt_c\partial^b_r -
  \frac{rl^2}{(ml^2-r^2)^2}\partial_tdr_c\partial_t^b +
  \frac{r}{l^2}\partial_r^ddt_c\partial_t^b -
  -\frac{r}{l^2}\partial_r^ddr_c\partial_r^b +
  \frac{ml^2+l^2-r^2}{rl^2}\partial_{\varphi}^d d\varphi_c\partial_r^b \right)
\end{eqnarray}
Note that the right-hand side of (\ref{Rdc}) is the derivative of
a locally integrable tensor. Therefore (\ref{Rdc}) defines a
distribution.

An analogous calculations to that of (\ref{ricdis}) leads to
\begin{equation}\label{eq26}
  R^a_{\, b} = \pi (1+m) \delta^{(2)}_{(0)} (\partial_x^a dx_b +
  \partial_y^a dy_b) - \frac{2}{l^2} (\partial_t^a t_b +
  \partial_x^a dx_a + \partial_y^a dy_b)
\end{equation}

Finally, from (\ref{chrissc},\ref{doschrissc}) and
(\ref{ERM},\ref{eq26}) we obtain
\begin{equation}\label{eq27}
  G^a_{\ b} - \frac{1}{l^2} (\partial_t^a t_b +
  \partial_x^a dx_a + \partial_y^a dy_b) = -\pi (1+m)
  \delta^{(2)}_{(0)} \partial_t^a dt_b.
\end{equation}
Remarkably enough, the right-hand side of (\ref{eq27}) resembles
the physically expected result for the distributional energy
momentum tensor $T^a_{\ b} = -m\delta^{(3)}_{(0)}\partial_t^a
dt_b$ of the Schwarzschild four dimensional black hole
\cite{parker,balasin93,balasin94,pantoja,heinzle}.\footnote{The
Schwarzschild metric is not a semi-regular metric and cannot be
handled with the methods used here to obtain its distributional
curvature\cite{pantoja}.}

Now, let us consider the dependence of these results on the
coordinate system. Note that whether or not a metric is
semi-regular depends in general on the differentiable structure
imposed on the manifold. In this section, the choice of the
manifold differentiable structure was made on the basis of an
interpretation of the coordinate system in which the metric is
given: we use Cartesian coordinates associated with the
Schwarzschild coordinates. In the next section the distributional
curvature and Einstein tensor fields are evaluated using the
Kerr-Schild form of the BTZ metric. This amounts to change both
the coordinates and the differentiable structure.

\section{BTZ black hole in Kerr-Schild
coordinates}\label{sec3}

In previous works the Kerr-Schild form of the Schwarzschild metric
has been proved to be useful, from both conceptual and technical
points of view, for the analysis of the distributional
Schwarzschild geometry from quite different approaches
\cite{parker,balasin94,heinzle}. In the following we shall prove
that the non-rotating BTZ solution in Kerr-Schild coordinates is a
semi-regular metric.

The $AdS_3$ black hole solution of BTZ is given in the Kerr-Schild
form by \cite{kim}
\begin{equation}
g_{ab}= \eta_{ab} + (1+m-\frac{r^2}{l^2})k_a k_b, \label{btzks}
\end{equation}
where $r=\sqrt{x^2+y^2}$ and
\begin{equation}
k_a= dt_a + \frac{1}{r}(xdx_a +ydy_a),
\end{equation}
with $k^a= \eta^{ab}k_b$ a null vector field with respect to
$\eta_{ab}$ and $g_{ab}$. It follows that there are two metrical
structures, $\eta_{ab}$ and $g_{ab}$, associated to the manifold.
We choose as the underlying manifold structure that of ${\cal
R}^3$ with the smooth metric $\eta_{ab}$ in Cartesian coordinates
$\{t,x,y\}$. Note that in Kerr-Schild coordinates $r$ is a
spacelike coordinate $\forall r> 0$, which is not the case in
Schwarzschild type coordinates. Note also that the natural volume
element $\omega_{g}$ associated to $g_{ab}$ agrees with the volume
element $\omega_{\eta}$ of $\eta_{ab}$.

Now,
\begin{equation}
(g^{-1})^{ab}= \eta^{ab} - (1+m-\frac{r^2}{l^2})\left(
\partial_t^a-\frac{1}{r}(x\partial_x^a+y\partial_y^a)\right)
\left(
\partial_t^b-\frac{1}{r}(x\partial_x^b+y\partial_y^b)\right).
\end{equation}
Clearly, $g_{ab}$ and $(g^{-1})^{ab}$ are locally integrable
$\forall m$ (Actually, $g_{ab}$ and $(g^{-1})^{ab}$ are locally
bounded).

The weak derivative in $\eta_{ab}$ of $g_{ab}$ exists almost
everywhere and is given by
\begin{equation}
 \nabla_c g_{ab} [U^{cab}] = \int_{{\cal R}^3} W_{cab}
 U^{cab} \omega_\eta,
\end{equation}
where
\begin{equation}
W_{cab}= -\frac{r}{l^2}dr_c(dt_a+dr_a)(dt_b+dr_b) +
\frac{1}{r}(1+m-\frac{r^2}{l^2})rd\varphi_c
\left(rd\varphi_a(dt_b+dr_b)+(dt_a+dr_a)rd\varphi_b \right),
\end{equation}
which is locally integrable $\forall m$.

From (\ref{chris}), it follows
\begin{eqnarray}\label{chrisks}
C^c_{ab} &=&
\frac{2r}{l^2}\left(dr_a(dt_b+dr_b)+dr_b(dt_a+dr_a)\right)(\partial_t^c-\partial_r^c)
+\frac{2r}{l^2}(1+m-\frac{r^2}{l^2})(dt_a+dr_a)(dt_b+dr_b)(\partial_t^c-\partial_r^c)
\nonumber\\ && +\frac{2r}{l^2}(dt_a+dr_a)(dt_b+dr_b)\partial_r^c
-\frac{2}{r}(1+m-\frac{r^2}{l^2})r^2d\varphi_ad\varphi_b(\partial_t^c-\partial_r^c),
\end{eqnarray}
which is locally integrable $\forall m$. Note that $C^b_{ab}=0$.

Finally, from (\ref{chrisks}) we find
\begin{eqnarray}
 2C^d_{m[b}C^m_{a]c}&=&2\frac{r^2}{l^4}(dt_adr_b-dr_adt_b)(dt_c+dr_c)(\partial_t^d-\partial_r^d)
 \nonumber\\
 &&-\frac{1}{l^2}(1+m-\frac{r^2}{l^2})\left((dt_a+dr_a)rd\varphi_b-rd\varphi_a(dt_b+dr_b)\right)
 rd\varphi_c(\partial_t^d-\partial_r^d),
\end{eqnarray}
which is locally integrable $\forall m$. Therefore the metric
(\ref{btzks}) is semi-regular. Furthermore, since (\ref{btzks}) is
a semi-regular metric $\forall m$, we can now consider the
distributional geometry of the BTZ black hole including the $m=0$
black vacuum.

We now calculate the Ricci tensor of (\ref{btzks}). We find
\begin{equation}
R_{ac}[U^{ac}]= -\int_{{\cal R}^3} C^b_{ac}\nabla_b U^{ac}
\omega_{\eta} - \int_{{\cal
R}^3}\frac{2r}{l^2}(dt_a+dr_a)(dt_c+dr_c)U^{ac}\omega_{\eta}.
\end{equation}
An analogous calculation to that of (\ref{ricdis}) leads to
\begin{equation}\label{ricdisks}
  R_{ac} = \pi(1+m) \delta^{(2)}_{(0)} (dx_a dx_c + dy_a dy_c) -
  \frac{2}{l^2} g_{ac},
\end{equation}
where $g_{ac}$ is the locally integrable tensor (\ref{btzks}).
Note that (\ref{ricdisks}) is equivalent to (\ref{ricdis}) which
was obtained using Schwarzschild type coordinates.

Calculations analogous to the ones done previously show that
\begin{equation}\label{eq26ks}
  R^a_{\ b} = \pi (1+m) \delta^{(2)}_{(0)} (\partial_x^a dx_b +
  \partial_y^a dy_b) - \frac{2}{l^2} (\partial_t^a t_b +
  \partial_x^a dx_a + \partial_y^a dy_b)
\end{equation}
and
\begin{equation}\label{eq27ks}
  G^a_{\ b} - \frac{1}{l^2} (\partial_t^a t_b +
  \partial_x^a dx_a + \partial_y^a dy_b) = -\pi (1+m)
  \delta^{(2)}_{(0)} \partial_t^a dt_b.
\end{equation}

\section{Discussions}\label{sec4}

As noted above, the distributional Ricci tensor fields
(\ref{ricdis}) and (\ref{ricdisks}) are equivalent, as is the case
with the mixed index versions (\ref{eq26}) and (\ref{eq26ks}). We
take (\ref{ricdisks}) and (\ref{eq26ks}), which are valid $\forall
m$, as the distributional $R_{ab}$ and $R^a_{\ b}$ Ricci tensor
fields. The non-rotating BTZ black hole geometry is singular and
this singularity is a curvature singularity proportional to a
$\delta$-distribution supported at the origin. As follows from
(\ref{ricdisks}) and (\ref{eq26ks}), even for the $m=0$ black
vacuum this singularity is present. As expected, for $m=-1$
(anti-de Sitter spacetime) a non-singular spacetime is recovered.

On the other hand, (\ref{eq27}) and (\ref{eq27ks}) are equivalent.
Hence, we take (\ref{eq27ks}) as the distributional Einstein
tensor field $G^a_{\ b}$ of the non-rotating BTZ black hole. As
expected on physical grounds, the distributional energy momentum
tensor $T^a_{\ b}$ of the BTZ black hole geometry is then given by
\begin{equation}\label{T}
T^a_{\ b}= -(1+m) \delta^{(2)}_{(0)}\partial_t^a dt_b,
\end{equation}
where we have set the gravitational constant $G$ equal to
$\frac{1}{8}$. Note that (\ref{T}) is a non-zero distribution for
$m=0$. The constant shift in the mass is due to the fact that in
(\ref{btzsc}) the zero point of energy has been set so that the
mass vanishes when the horizon size, the length of the minimal
geodesic of the horizon, goes to zero \cite{banadosPRD}. As
follows from the fact that the singular parts of the
distributional curvature and Einstein tensor fields are equal to
zero only for $m=-1$, if the zero of energy is adjusted so that
anti-de Sitter space has zero mass then $1+m\rightarrow m$.

Finally, let us briefly discuss the coordinate independence of
these results. As already mentioned in section \ref{sec2}, whether
or not a given metric is semi-regular depends in general on the
differentiable structure imposed on the underlying manifold. In
sections \ref{sec2} and \ref{sec3}, the choice of the
differentiable structure was made on the basis of an
interpretation of the coordinate system in which the metric is
given. In this sense, the choice of coordinates determines the
differentiable structure of the underlying manifold.  Here we find
that the distributional curvature and Einstein tensor fields are
well defined and that they are equivalent in both descriptions.
Thus, in a restricted sense, the present distributional treatment
in Schwarzschild and Kerr-Schild coordinates provides invariant
results for the distributional curvature and Einstein tensor
fields of the BTZ black hole geometry.

\section*{Acknowledgements}

We wish to thank Jorge Zanelli for fruitful discussions. This work
was financed by CDCHT-ULA under project C-1073-01-05-A.

\end{document}